\documentclass[aps, prl, twocolumn, superscriptaddress, longbibliography,
]{revtex4-2}

\usepackage{graphicx}
\usepackage{natbib, hyperref}

\newcommand*{\reffigure}[2]{\ref{#1}(#2)}

\begin{document}

\title{Universal transparency and fine band structure near the Dirac point in HgTe quantum wells}

\author{V.~Dziom}
\email[E-mail:~]{udziom@ist.ac.at}
\affiliation{Institute of Science and Technology Austria,
3400 Klosterneuburg, Austria}
\author{A.~Shuvaev}
\author{J.~Gospodari\v{c}}
\affiliation{Institute of Solid State Physics,
Vienna University of Technology, 1040 Vienna, Austria}
\author{E.~G.~Novik}
\affiliation{Institute of Theoretical Physics, Technische Universit\"{a}t
Dresden, 01062 Dresden, Germany}
\author{A.~A.~Dobretsova}
\author{N.~N.~Mikhailov}
\author{Z.~D.~Kvon}
\affiliation{Rzhanov Institute of Semiconductor Physics, 630090 Novosibirsk,
Russia}
\affiliation{Novosibirsk State University, 630090 Novosibirsk, Russia}
\author{Z.~Alpichshev}
\affiliation{Institute of Science and Technology Austria,
3400 Klosterneuburg, Austria}
\author{A.~Pimenov}
\affiliation{Institute of Solid State Physics,
Vienna University of Technology, 1040 Vienna, Austria}

\begin{abstract}%
Spin--orbit coupling in thin HgTe quantum wells results in a relativistic-like electron band structure, making it a versatile solid state platform to observe and control non-trivial electrodynamic phenomena. Here we report an observation of universal terahertz (THz) transparency determined by fine-structure constant $\alpha \approx 1/137$ in 6.5\,nm-thick HgTe layer, close to the critical thickness separating phases with topologically different electronic band structure. Using THz spectroscopy in magnetic field we obtain direct evidence of asymmetric spin splitting of the Dirac cone. This particle-hole asymmetry facilitates optical control of edge spin currents in the quantum wells.%
\end{abstract}

\date{\today}
\maketitle
While advance in fundamental high-energy physics requires increasingly larger and more expensive facilities, solid crystals provide an alternative platform allowing to explore relativistic electrodynamics. Interaction of an electron with the crystal field leads to a modification of the energy vs momentum relation. In particular cases it results into a relativistic-like dispersion, the most well-known example being graphene. The relativistic spectrum leads to various unusual electrodynamic phenomena \cite{Ren_Nat_2019,Muller_NanoLetters_2021,Mahler_PRX_2019,Shamim_NatComm_2021,Strunz_NatPhys_2020,wu_science_2016,dziom_ncomm_2017,Kadykov_PRL_2018,But_NatPh_2019,Piatrusha_PRL_2019,park_ncomms_2015,armitage_rmp_2018,basov_rmp_2014}. 
The highest-quality graphene is currently available in the form of microscopic flakes and great efforts are made to improve techniques for production of macroscopic sheets. Demand for larger defect-free samples is coming not only from practical applications, but also from fundamental research, because usage of numerous powerful experimental methods is impossible or strongly limited for microscopic samples.

Two-dimensional electronic states with relativistic-like spectrum can also be realized in HgTe/CdHgTe heterostructures, grown using molecular beam epitaxy. Constant improvement in the technology of growth \cite{Varavin_OIDP_2020} and creation of semi-transparent gate made it possible to use magneto-optical terahertz (THz) spectroscopy to study fine details of the electronic spectrum around the Dirac point.
In present work we report the observation of universal transmission of coin-sized HgTe/CdHgTe heterostructures defined by the fundamental quantum fine-structure constant $\alpha=e^2/(4 \pi \varepsilon_0 \hbar c) \approx 1/137$.
Unlike the optical quantum Hall effect (QHE) \cite{wu_science_2016,shuvaev_prl_2016,dziom_ncomm_2017,dziom_prb_2019} and similar to the anomalous QHE \cite{okada_ncomm_2016}, in the reported case the universal absorption $\pi \alpha/2 \approx 1.1\%$ is observed in zero magnetic field, while the application of magnetic field only facilitates the measurement of this tiny effect.
Also, our study of the cyclotron resonance provides direct evidence of significant electron-hole asymmetry at the Dirac point, that we attribute to much stronger spin splitting in the valence band. Using the semi-classical approach we reconstruct fine details of the band structure around the Dirac point. The particle-hole asymmetry, that can be useful in quantum devices like nano-transistors \cite{McRae_ncomm_2017} and magnetically tunable lasers \cite{But_NatPh_2019}, has been long overlooked in theoretical models for CdHgTe heterostructures. Recent theoretical works \cite{Kaladzhyan_PRB_2015,mahmoodian_jpcm_2017,durnev_annalen_2019,durnev_jpcm_2019,durnev_prb_2021} suggest that particle-hole asymmetries of bulk states enable to photo-induce spin-polarized DC currents in the helical edge states of the CdHgTe quantum wells.
\begin{figure*}[tbp]
\centerline{\includegraphics[width=1.00\linewidth,clip]{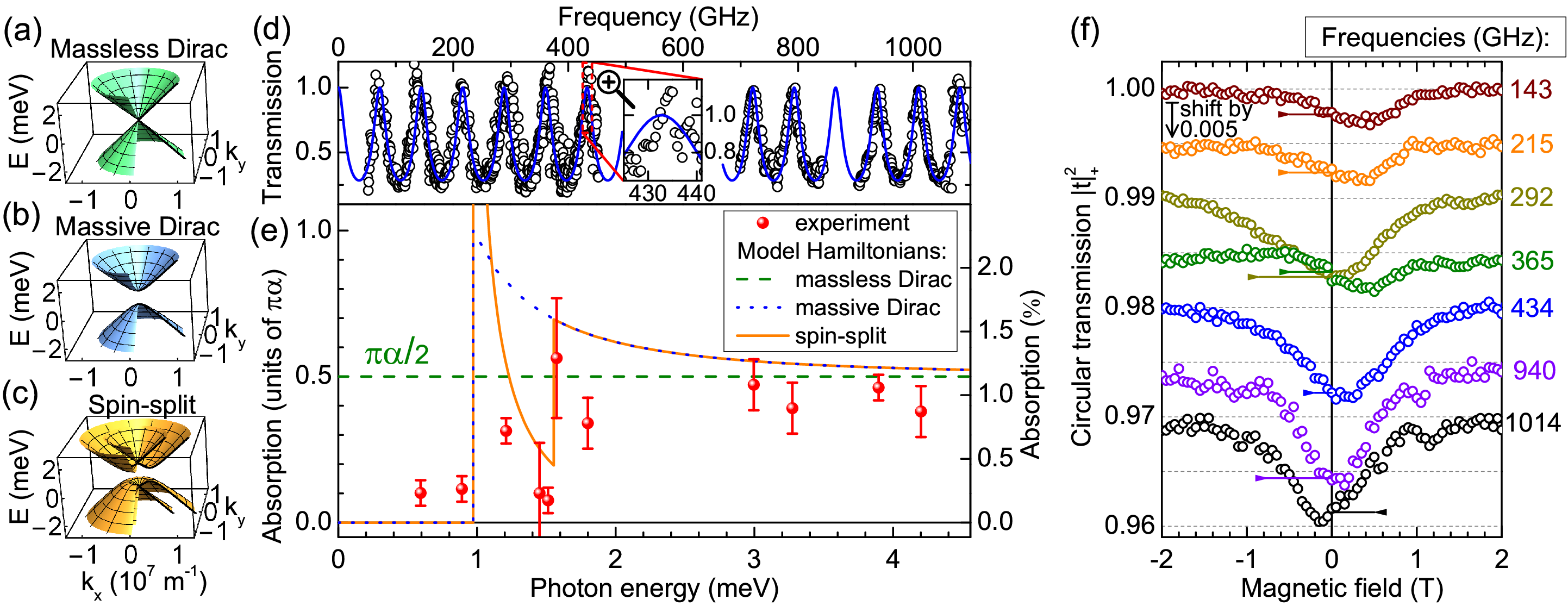}}
\caption{
Observation of absorption defined by the fine-structure constant $\alpha$ in macroscopic HgTe/CdHgTe heterostructures.
(a--c) Electronic band structure in several theoretical models.
(d) Directly measured transmission coefficient for THz radiation in zero field at 1.85\,K.
(e) Absorption in zero field
as extracted from field dependence of transmission at fixed frequencies (f).
}
\label{figure1}
\end{figure*}

Electronic band structure in the family of cadmium-mercury telluride (HgTe/Cd$_{1-x}$Hg$_{x}$Te)
quantum wells can be engineered by growth procedure~\cite{bernevig_science_2006, buttner_nphys_2011, kvon_jetpl_2012}.
The energy gap strongly depends on the thickness of HgTe layer $d$ and weakly on the barrier composition $x$. 
HgTe wells with thickness around $d_c=6.5$\,nm demonstrate vanishing gap in transport measurements.
Interest to these quantum wells was greatly enhanced when the calculation of the band structure using $\mathbf{k} \cdot \mathbf{p}$ model~\cite{bernevig_science_2006} predicted a topological transition with the relativistic electron spectrum
\begin{equation}
E(\mathbf{k})=\pm \sqrt{\delta^2+(v_F \hbar \mathbf{k})^2},
\label{eqDirac}
\end{equation}
where the band gap $2|\delta| \propto |d_c-d|$ in the vicinity of $d_c$ and turns to zero at $d=d_c$. In this simple model at $d=d_c$ the bands form an axial-symmetric spin-degenerate cone shown in
Fig.~\reffigure{figure1}{a}, 
corresponding to a massless relativistic Dirac quasiparticle. If the Fermi level is tuned to the Dirac point, e.g. the lower half of the cone is filled, and a plane electromagnetic wave passes through the 2D system, the electric field will induce vertical electron transitions between the valence and conduction bands, leading to energy absorption rate $\pi \alpha/2 \approx 1.1\%$, shown in
Fig.~\reffigure{figure1}{e}
by green dashed line. The value of absorption does not depend on the material parameter $v_F$ in a range of photon energies, thus demonstrating so called universal behavior. In graphene \cite{nair_science_2008,buttner_nphys_2011} the two-fold valley degeneracy leads to the twice higher absorption equal to $\pi \alpha$. Energy-independence of the absorption allows a simple intuitive explanation based on dimensional analysis \cite{mak_ssc_2012}: the linear spectrum $E=v_F \hbar k$ doesn't contain any energy scales to form a dimensionless combination with the photon energy for dimensionless absorption $\eta(\hbar \omega)$.

A realistic HgTe quantum well consists of an integer number of atomic layers and its thickness cannot exactly match the theoretical value.
Samples with nominally critical thickness can demonstrate small band gaps of the order of several meV.
Opening of the gap in spectrum~(\ref{eqDirac}) leads to modification of absorption \cite{bernevig_science_2006,nair_science_2008,tarasenko_prb_2015} at photon energies $\hbar \omega \lesssim 2 \delta$ as shown in Fig.~\reffigure{figure1}{e} by blue dotted line. The quantum well does not absorb radiation with the photon energy $\hbar \omega<2 \delta$, and just above the band gap the absorption is doubled. At photon energies much higher than the gap $\hbar \omega \gg 2 \delta$
absorption tends to the gap-independent value $\eta_0=\pi \alpha /2$.

Derivation of the seminal model~\cite{bernevig_science_2006} for the Dirac spectrum~(\ref{eqDirac}) used simplifications that led to spin-degenerate solutions for electron states. Later the model was modified~\cite{Liu_prl_2008,orlita_prb_2011,minkov_prb_2014,tarasenko_prb_2015,minkov_prb_2016} to allow for the bulk inversion asymmetry of host CdHgTe and HgTe crystals and the inversion asymmetry of their interface. In Ref.~[\onlinecite{tarasenko_prb_2015}] the modified effective Hamiltonian has a solution in the form of spin-split electron-hole symmetric bands
\begin{equation}
E(\mathbf{k})=\mp \sqrt{\delta^2+(v_F \hbar k \pm \gamma )^2},
\label{spinSplitEnergy}
\end{equation}
shown in Fig~\reffigure{figure1}{c}. The electron states and matrix elements which determine absorption are also modified, resulting in an absorption peak at $\hbar \omega = 2 \delta$ and a local minimum in absorption just below $\hbar \omega=2\sqrt{\delta^2+\gamma^2}$. An example of absorption calculated within this model is shown as solid orange line in Fig.~\reffigure{figure1}{e} with the parameters $2\delta=0.97$\,meV and $2\gamma=1.21$\,meV giving the optimal fit to the experimentally determined absorption in 6.5\,nm HgTe quantum well.

The experiments reported here were initially intended for reconstruction of the band structure by measuring the cyclotron resonances in perpendicular magnetic field at sub-THz frequencies. 
In such experiment continuous circularly- or linearly-polarized radiation passes through a heterostructure at T=1.85\,K and the absolute value of its transmission coefficient is determined as the ratio of intensities with and without the sample in the beam: $|t|^2=I_s/I_0$. 
In order to determine the complex dynamic conductivity $\sigma_{xx}=\sigma_{xx}'+\imath \sigma_{xx}''$ of the quantum well at an arbitrary frequency, one generally needs the complex phase $\arg t$ as well.
The estimated phase shift due to the interband transitions is an order of magnitude smaller, than $\sim 1$\,$\mu$m reproducibility of our Mach-Zehnder interferometer.
Fortunately, the corresponding Fresnel equations are greatly simplified in Fabry-P\'{e}rot transmission maxima.
At these frequencies the substrate effectively disappears from the equations and the sample behaves as a free-standing HgTe well, so that the phase in not needed to determine absorption.
If the Fermi level is tuned to the neutrality point, then electrodynamics reduces to
\begin{equation}
\eta=\sigma_{xx}' Z_0= 1-|t|^2,
\end{equation}
where $Z_0\approx 377$\,$\Omega$ is the impedance of free space

Direct accurate measurements of transmission in an optical cryostat are complicated by reflections between the sample and the optical windows, which are hard to account for analytically. These reflections introduce a reproducible oscillatory deviation that can be seen in Fig.~\reffigure{figure1}{d}, where symbols show experimental transmission through the 6.5\,nm well and the blue curve is a calculation for a bare substrate in vacuum.
In order to extract the small absorption in the well, we measured circular or linear transmission coefficient at fixed frequencies as a function of a perpendicular magnetic field, see Fig.~\reffigure{figure1}{f}.
Using an analytical transfer-matrix method and finite-element numerical simulations~\cite{dziom_phd_2018} we verified that at a fixed frequency the unaccounted reflections mostly act as a ``multiplier'', preserving the relative changes in transmission.
In the applied field of several Tesla separation between Landau levels exceeds the photon energy and the well becomes fully transparent.
Thus we normalize the experimental data by the high-field value to recover transmission in free space and absorption in zero field, as shown in Fig.~\reffigure{figure1}{e,f}. This procedure can be used with both circular and linear polarizations, because in zero field the transmission coefficients are equal: $t_{||}|_{B=0}=t_{\pm}|_{B=0}$.

Experimental absorption is shown in Fig.~\reffigure{figure1}{e} as a function of frequency.
For two points adjacent to 365\,GHz that are out of a Fabry-P\'{e}rot maximum, absorption in HgTe well was recalculated using the phase from spin-split model calculation, shown by the orange curve. At high frequencies $\hbar \omega>1.5$\,meV experimental absorption
tends to the universal value $\pi \alpha /2$, common for all the models discussed above and at low frequencies the well is almost transparent. The onset of significant absorption occurs at photon energies around $1$\,meV and we take $E_g = 2 \delta= 0.97$\,meV from the spin-split model as an experimental estimate of the band gap to reconstruct the band structure as described further in the text.

\begin{figure*}[tbp]
\centerline{\includegraphics[width=1.00\linewidth,clip]{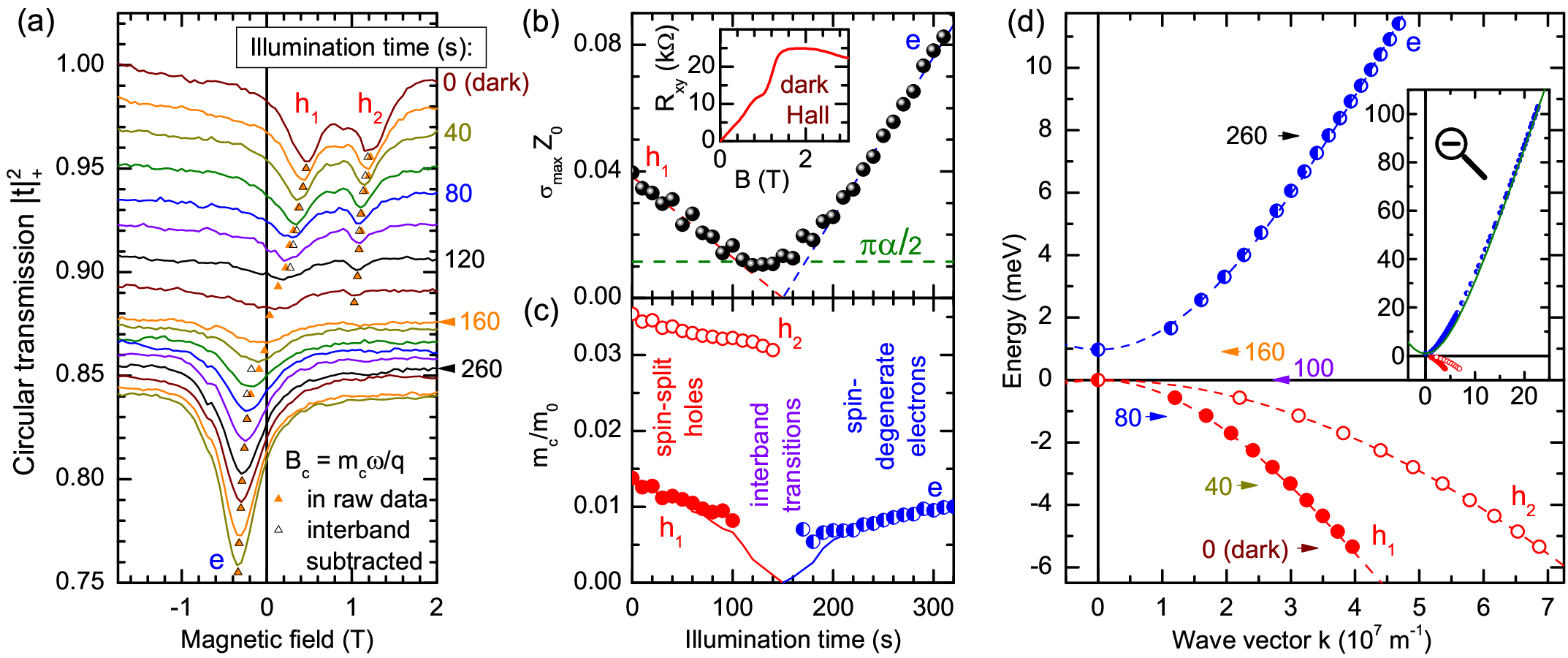}}
\caption{
Using cyclotron resonances to reconstruct fine details of the electronic band structure. (a) Transmission of circularly polarized 940\,GHz radiation at 1.85\,K. Dark sample demonstrates two resonances at positive magnetic field, corresponding to spin-resolved hole subbands. Illumination shifts the Fermi level to the conduction band with much smaller spin splitting, and electrons produce the single resonance in the negative field. (b) Optical conductivity in resonances $h_1$ and $e$ takes the universal value $\pi \alpha/2$ upon passing through zero magnetic field. $Z_0$ is the vacuum impedance; dashed lines are guides to the eye. The inset shows in situ Hall resistance for the dark sample. (c) Cyclotron masses determined from positions of the resonances after subtraction of the interband absorption (symbols) and in the raw data (solid lines). (d) Reconstructed electronic band structure assuming axial symmetry. Numbers with arrows indicate illumination time for the corresponding energy level.
}
\label{figure2}
\end{figure*}
At high energies the 6.5\,nm HgTe well demonstrates universal behavior that is insensitive to fine details of the band structure near the charge neutrality point. However at $\hbar \omega \simeq E_g$ absorption is in quantitative disagreement with the electron-hole-symmetric models discussed above. This discrepancy may be an indication of an asymmetry between conduction and valence bands. An electron--hole asymmetry (formation of additional hole pockets) at energies below $\simeq 20$\,meV is predicted by $\mathbf{k} \cdot \mathbf{p}$ calculations and confirmed experimentally on wells of similar thicknesses \cite{buttner_nphys_2011,Hubmann_JIMTW_2020,
minkov_prb_2014,Kuntsevich_PRB_2020,gospodaric_prb_2020}. In our cyclotron resonance (CR) experiments we clearly observe asymmetry between the conduction (CB) and valence (VB) bands even at $E \simeq 1$\,meV. 

In a CR experiment we sweep the magnetic field while the frequency and the Fermi level are fixed.
Between the sweeps the Fermi level can be tuned by the effect of persistent photoconductivity or by a semi-transparent gate.
Here we mainly focus on the experiments with the optical doping.

Upon cooling down sample~\#1 in darkness we obtain the Fermi level in VB, with the density of holes $n_{\text{Hall}}=5\times10^{10}$\,cm$^{-2}$.
This value, determined by intrinsic doping, was obtained by measuring $R_{H}$ slope of the Hall resistance, shown in the inset in Fig.~\reffigure{figure2}{b}.
By illuminating the sample with a green LED diode we shifted the Fermi level by small steps through the charge neutrality point to CB. 
Figure~\reffigure{figure2}{a} shows transmission coefficient $|t_+|^2$ of circularly polarized light at $940$\,GHz as a function of perpendicular magnetic field $B$ (in Faraday geometry). Each curve was taken at a fixed value of the Fermi level by keeping the sample in darkness during each sweep of magnetic field and illuminating it only between the successive sweeps. In such experiment, quasiparticles with a parabolic band dispersion $E(k)=\pm (\hbar k)^2/(2 m)$ would produce a single dip at either positive (for holes) or negative (for electrons) magnetic field $B_c=m \omega/e$. In sample~\#1
the initial Fermi level lies in VB and the transmission demonstrates two absorption resonances in positive fields, corresponding to two sorts of quasiparticles (hole subbands). As the Fermi level is shifted to the top of VB, the amplitude of both resonances decreases and the low-field resonance $h_1$ approaches zero field.
After 150\,s of total illumination resonance $h_1$ reaches its minimal amplitude and passes through $B=0$, while resonance $h_2$ almost vanishes at this point. The study of the universal transparency at different frequencies (Fig.~\reffigure{figure1}{e}) was conducted at this position of the Fermi level. With further illumination the resonance at $B=0$ increases and shifts to the region $B<0$, corresponding to negative quasiparticles.
Figure~\reffigure{figure2}{b} shows the value of $\sigma'_{xx}$, corresponding to the maximal absorption in resonances $h_1$--e. Remarkably, absorption saturates at the universal value of $\pi \alpha /2$ around the charge neutrality point. Outside this region the conductivity changes linearly with illumination up to 320\,s, where it starts showing signs of saturation.

The most probable explanation of e--h asymmetry in the data in Fig.~\reffigure{figure2}{a} is a much stronger spin splitting in VB in comparison to CB. Evidences of such difference were demonstrated in SdH transport experiments in tilted magnetic field, supported by theoretical calculations using the tight-binding method \cite{minkov_prb_2016}. Following this interpretation, we attribute hole resonances $h_1$ and $h_2$ to spin-resolved CR in VB, whereas CR in CB appears as the single peak due to the smaller splitting.
Analysis of temperature-dependent SdH oscillations in a 6.0\,nm HgTe well \cite{Kuntsevich_PRB_2020} provided estimates for the masses of spin-split subbands $m_1/m_0=0.015$ and $m_2/m_0=0.03$, that are very close to the directly measured cyclotron masses in Fig.~\reffigure{figure2}{c}. We note that we were unable to find convincing spectroscopic evidences of heavy-hole pockets \cite{buttner_nphys_2011,Hubmann_JIMTW_2020,minkov_prb_2014,minkov_prb_2016,Kuntsevich_PRB_2020,gospodaric_prb_2020} with $m_c/m_0 \sim 0.2$ in the samples of critical thickness, even upon an application of extreme negative gate voltages. Due to the much greater expected mass the corresponding CR peak strongly broadens and becomes hard to be detected at low densities.

While an appropriate model is to be developed for a consistent quantum description of electron states and transitions in magnetic field, we use the semiclassical approach which allows to reconstruct the approximate band structure and to estimate the value of spin splitting in VB. The semiclassical description for electron conductivity in magnetic field requires several conditions to be fulfilled, among which are: (I) electrons fill many Landau levels (filling factor $\nu \gg 1$) and (II) the interband transitions can be neglected. Both conditions are violated when the Fermi level approaches the gap. As the photon energy we use ($f=940\text{\,GHz}\approx4$\,meV) is greater than the measured band gap (1\,meV), the interband transitions produce additional absorption in zero field. It could be avoided by using a photon energy $\hbar \omega<1$\,meV, but unfortunately the hole resonances are not resolved as separate peaks at such small frequencies, that prevented their observation in our earlier work \cite{shuvaev_prb_2017}. To extrapolate the approach to the small densities, we subtract the zero-field resonance absorption at the neutral point before using the semiclassical analysis. The limitation (I) is mitigated by using the cyclotron mass from spectroscopy in combination with density from in situ transport Hall measurements for the band reconstruction.

In order to reconstruct the band structure we use the method described in Refs.~[\onlinecite{shuvaev_prb_2017,gospodaric_prb_2020,gospodaric_prb_2021}], modified for the presence of two carriers simultaneously. We assume an axial symmetry $E_{h1,h2,e}(k_x,k_y)=\epsilon_{h1,h2,e}(k)$ and zero splitting \cite{Samokhin_AoP_2009} at the $\Gamma$-point: $\epsilon_{h1}(0)=\epsilon_{h2}(0)=0$.
After a short illumination at step $j$, the Fermi energy is shifted by $\Delta E_j$ and the density of a subband is changed by $\Delta n_j$. The definition of the cyclotron mass \cite{ashcroft_book} allows to determine $\Delta E_j$ as
\begin{equation}
\Delta E_j = \frac{2\pi \hbar^2}{D} \frac{\Delta n_j}{m_{c,j}},
\label{reconstructEnergy}
\end{equation}
with spin degeneracy $D=1$ for holes and $D=2$ for electrons.
Using this approach we obtain the band structure as shown in Fig.~\reffigure{figure2}{d}. Symbols show the results of the band reconstruction and dashed lines are guides to the eye, plotted using relativistic Dirac dispersion~(\ref{eqDirac}) for each subband independently. The data above 18\,meV was obtained in sample~\#3 with a semi-transparent gate, where the Fermi level could be tuned to $+100$\,meV.
Zero gate voltage corresponded to $E_F\approx 50$\,meV, thus the available range of energies laid mostly in CB. The value of the band gap $E_g=1$\,meV was determined using frequency-dependent absorption in Fig.~\reffigure{figure1}{e}.
The proximity of the hole subband $h_2$ to a perfect parabola is likely specific for 6.5\,nm wells, because the hole dispersion in the 7.1\,nm sample~\#4 was found to be strongly non-parabolic for both subbands. This fact implies that the growth procedure can be used to control
the spin-splitting and the fine band structure in VB.

Effects of a particle--hole asymmetry of bulk states on properties of helical edge states were recently analyzed theoretically \cite{Kaladzhyan_PRB_2015,mahmoodian_jpcm_2017,durnev_annalen_2019,durnev_jpcm_2019,durnev_prb_2021}. A bulk asymmetry can distort the dispersion of the edge states, which is linear in the symmetric case, and shift their charge neutrality point away from the middle of the bulk gap in nearly critical wells \cite{entin_epl_2017,durnev_annalen_2019}. Particle-hole asymmetry allows electron transitions between the edge and bulk states \cite{Kaladzhyan_PRB_2015,durnev_prb_2021}. Under incident light the transitions cause spin-polarized DC current in edge channels. Thus the asymmetry facilitates generation and optical control of spin currents using CdHgTe quantum wells.

Among 2D materials with the quasi-relativistic spectrum HgTe wells stand out due to development of MBE growth that made them available in the form of large high-quality wafers, suitable for both technological applications and fundamental research.
The large size allows to comprehensively study complex emergent phenomena by combining low-energy spectroscopy, transport, capacitance, thermodynamic and other techniques in situ on the same sample.
The observation of absorption determined by the fine-structure constant in coin-sized samples represents a phenomenon that is consistent with expectations for a Dirac spectrum. The precise knowledge of fine asymmetry is important for interpretation of any phenomena involving electronic properties near the charge neutrality point.
The CR technique that we used can be universally applied to 2D materials to determine fine details of their band structure.
%

\end{document}